\documentclass[12pt]{article}             
\usepackage{feynmf}
\newcommand{\note}[1]{}
\newcommand{\seclabel}[1]{\label{#1}}
\newcommand{\figlabel}[1]{\label{#1}}
\newcommand{\bib}[1]{\bibitem{#1}}

\newcommand{\3}{\ss}
\newcommand{\absatz}{\vspace{2ex}\noindent}

\newcommand{\journal}[4]{{#1} {\bf{#2}}, #3 (#4)}
\newcommand{\NPB}{\emph{Nucl.\ Phys.\ }{B}}
\newcommand{\PLB}{\emph{Phys.\  Lett.\ }{B}}
\newcommand{\PRD}{\emph{Phys.\ Rev.\ }{D}}
\newcommand{\book}[3]{\emph{#1}; #2 (#3)}
\newcommand{\preprint}[3]{\emph{#1}; #2, #3 (to be published)}

\newcommand{\dis}{\displaystyle}
\newcommand{\non}{\nonumber}

\newcommand{\half}{\frac{1}{2}}
\newcommand{\e}{\mathrm{e}}
\newcommand{\ii}{\mathrm{i}}
\newcommand{\dd}{\mathrm{d}}

\newcommand{\xs}{\vec{X}_{\mathrm{s}}}
\newcommand{\ts}{T_{\mathrm{s}}}

\newcommand{\xu}{\vec{X}_{\mathrm{u}}}
\newcommand{\tu}{T_{\mathrm{u}}}
\newcommand{\Qs}{Q_{\mathrm{s}}}
\newcommand{\Qp}{Q_{\mathrm{p}}}
\newcommand{\calQs}{{\calQ_{{\rm s}}}}
\newcommand{\calQp}{{\calQ_{{\rm p}}}}

\newcommand{\Phis}{\Phi_{\mathrm{s}}}
\newcommand{\Phip}{\Phi_{\mathrm{p}}}

\newcommand{\As}{A_{\mathrm{s}}}
\newcommand{\Ap}{A_{\mathrm{p}}}
\newcommand{\Au}{A_{\mathrm{u}}}
\newcommand{\calAs}{\calA_{\mathrm{s}}}

\newcommand{\Amu}{{A^\mu}}
\newcommand{\Asmu}{{A_{{\rm s}}^\mu}}
\newcommand{\Apmu}{{A_{{\rm p}}^\mu}}
\newcommand{\Aumu}{{A_{{\rm u}}^\mu}}
\newcommand{\Av}{\vec{A}}
\newcommand{\Avs}{{\vec{A}_{{\rm s}}}}
\newcommand{\Avp}{{\vec{A}_{{\rm p}}}}
\newcommand{\Avu}{{\vec{A}_{{\rm u}}}}
\newcommand{\calAsmu}{{\calA_{{\rm s}}^\mu}}
\newcommand{\calApmu}{{\calA_{{\rm p}}^\mu}}
\newcommand{\calAumu}{{\calA_{{\rm u}}^\mu}}

\newcommand{\Xvs}{\vec{X}_{\mathrm{s}}}

\newcommand{\Ts}{T_{\mathrm{s}}}

\newcommand{\dedreiXs}{\dd^{3}\:\!\! X_{\mathrm{s}}\;}
\newcommand{\dedreiXu}{\dd^{3}\:\!\! X_{\mathrm{u}}\;}

\newcommand{\deTs}{\dd T_{\mathrm{s}}\;}
\newcommand{\deTu}{\dd T_{\mathrm{u}}\;}

\newcommand{\dedk}{\frac{\dd^{d}\! k}{(2\pi)^d}\;}

\newcommand{\deintdim}[2]{\frac{\dd^{#1}\! #2}{(2\pi)^{#1}}\;}

\newcommand{\kv}{\vec{k}}
\newcommand{\pv}{\vec{\,\!p}\!\:{}}
\newcommand{\qv}{\vec{q}}

\newcommand{\xv}{\vec{x}}

\newcommand{\de}{\partial}
\newcommand{\dev}{\vec{\de}}

\newcommand{\calA}{{\cal A}}\newcommand{\calC}{{\cal C}}

\newcommand{\calL}{{\cal L}}\newcommand{\calO}{{\cal O}}
\newcommand{\calQ}{{\cal Q}}

\begin{document}

\begin{fmffile}{calfeyn}
\fmfset{curly_len}{2mm}
\fmfset{wiggly_len}{3mm}
\newcommand{\feynbox}[2]{\mbox{\parbox{#1}{#2}}}
\newcommand{\fs}{\scriptstyle} 
\newcommand{\hq}{\hspace{0.5em}}
  
%

\begin{titlepage}
\begin{flushright}
hep-ph/9804251\\ NT@UW-98-12  \\ 1st April 1998 \\
\end{flushright}
\vspace*{1.5cm}
\begin{center}

\LARGE{\textbf{The Soft R\'egime in NRQCD}\footnote{Talk presented at the
    workshop ``Nuclear Physics with Effective Field Theories'' at Caltech,
    26th -- 27th February 1998.}}

\end{center}
\vspace*{1.0cm}
\begin{center}
\textbf{Harald W.\ Grie\3hammer\footnote{Email: hgrie@phys.washington.edu}}

\vspace*{0.2cm}

\emph{Nuclear Theory Group, Department of Physics, University of Washington,\\
Box 351 560, Seattle, WA 98195-1560, USA} \vspace*{0.2cm}

\end{center}

\vspace*{2.0cm}

\begin{abstract}
  A Lagrangean and a set of Feynman rules are presented for Non-Relativistic
  QCD with manifest power counting in the heavy quark velocity $v$. A r\'egime
  is identified in which energies and momenta are of order $Mv$. It is neither
  identical to the ultrasoft r\'egime corresponding to radiative processes with
  energies and momenta of order $Mv^2$, nor to the potential r\'egime with on
  shell quarks and gluons providing the Coulomb binding. In this soft r\'egime,
  gluons are on shell, and the quark propagator becomes static. Examples show
  that it contributes to one- and two-loop corrections of scattering and
  production amplitudes near threshold.  The results are readily generalised to
  any effective field theory with more than one low energy scale.
\end{abstract}
\vskip 1.0cm
\noindent
\end{titlepage}

\setcounter{page}{2}
\setcounter{footnote}{0}
\newpage

%

\section{Introduction}
\seclabel{intro}

Why a talk on Non-Relativistic QCD~\cite{CaswellLepage,BBL} in a workshop
dedicated to nuclear physics with effective field theories?  First, a few words
on the background: NRQCD (which in the following also includes NRQED) uses the
fact that systems consisting of two or more heavy quarks with mass $M$ become
very similar to the hydrogen atom: The Coulomb interaction rules the level
spacing in Charmonium and Bottomium because $\alpha_\mathrm{s}$ is small enough
for perturbative calculations. The relative velocity of the quarks in such
systems is $v\sim \alpha_\mathrm{s}(Mv)$ by virtue of the virial theorem, where
the scale at which the running coupling is to be taken is the inverse Bohr
radius $Mv$ of the system. Although $\alpha_\mathrm{s}$ increases with
decreasing $Q^2\sim (Mv)^2$, a window between the relativistic perturbative and
the confinement r\'egime remains in which both $\alpha_\mathrm{s}$ and $v$ is
small (for Bottomium, $\alpha_\mathrm{s}(Mv)\approx 0.2$). In the resulting
non-relativistic and perturbative framework, potentials and wave functions may
be used, so that bound state physics is easily accounted for, and calculations
of production cross sections, hyperfine splittings, lifetimes etc.\ are much
facilitated.

Like in effective nuclear theories, the NRQCD Lagrangean in terms of the heavy
quark (anti-quark) bi-spinors $Q$ ($\bar{Q}$) and gluons
($D_\mu=\de_\mu+igA_\mu$)
\begin{eqnarray}
  \lefteqn{\calL_\mathrm{NRQCD}=
        Q^\dagger\Big(\ii\de_0-gA_0\Big)Q \;+\;
        \frac{{c_1}}{2M}\;Q^\dagger \vec{D}^2 Q\;+
        \;\frac{{c_2}}{8M^3}\; Q^\dagger\vec{D}^4 Q\;+\dots}\non\\
      \label{nrqcdlagr}
    &&+ \;\frac{g{c_3}}{2M}\; Q^\dagger\vec{\sigma}\cdot\vec{B}Q\;+\dots
        \;-\;\frac{g^2{d_1}}{4M^2}\;\Big(\calC\bar{Q}^\dagger\vec{\sigma}Q\Big)
        \cdot\Big(Q^\dagger \vec{\sigma}\calC\bar{Q}\Big)\;+\dots\\
    &&-\;\frac{{e_1}}{4}\; F_{\mu\nu}^a F^{\mu\nu,\,a}+
        \;\frac{g^3{e_2}}{240 \pi^2 M^2}\;T(Q)F_{\mu\nu}^a  D^2
        F^{\mu\nu,\,a}+\;\dots+\;\calL_\mathrm{GFix}\non
\end{eqnarray}
consists of infinitely many terms constrained only by the symmetries of the
theory (here: gauge invariance) and is non-renormalisable. Predictive power is
nonetheless established when only a finite number of terms contribute to a
given order in the two expansion parameters\footnote{For clarity, the two will
  be distinguished in the following, although $v\sim\alpha_\mathrm{s}$.}, $g$
and $v$.  Excitations with four-momenta bigger than $M$ are integrated out,
giving rise e.g.\ to four-point interactions between quarks ($d_1\not=0$). So,
the ultraviolet physics is encoded in the coefficients $c_i,d_i,e_i$. An
advantage of NRQCD is that these can be determined by matching NRQCD matrix
elements to their QCD counterparts, as both are perturbative in the coupling
constant. At tree level, the Foldy-Wouthuysen transformation gives $c_i=1$, and
loop corrections are down by powers of $g$, the most famous example being the
coefficient for the Pauli term related to the anomalous magnetic moment of the
electron, $c_3=1+\frac{\alpha_\mathrm{s}}{2\pi}+\dots$. At one loop level,
further coefficients enter, and $d_i=e_i=1$. In contradistinction, the free
parameters of effective nuclear theories have to be determined from data
because it is still unclear how to match nuclear theory to QCD.

Another similarity between NRQCD and effective nuclear theory is the existence
of three scales: Besides the heavy quark mass $M$, the typical energy and
momentum scales in NRQCD are the bound state energy $Mv^2$ and the relative
momentum of the two quarks $Mv$ (i.e.\ the inverse size of the bound
state)~\cite{CaswellLepage,BBL}. In $NN$ scattering, the pion production
threshold $\sqrt{M_Nm_\pi}$ and the pion mass $m_\pi$ supplement the nucleon
mass $M_N$. As noted by Luke and Manohar~\cite{LukeManohar}, this allows one to
formally identify a small parameter in nuclear theories,
$\sqrt{\frac{m_\pi}{M_N}}\sim v$. In both theories, the effective Lagrangean
does not exhibit the non-relativistic expansion parameter explicitly, so that a
power counting scheme has to be established which determines uniquely which
terms in the Lagrangean must be taken into account to render consistent
calculations and predictive power to a given order in $v$. It is at this point
that NRQCD can serve as a ``toy model'' for effective nuclear theory (although
this grossly understates its value): It will establish what the relevant
kinematic r\'egimes and infrared variables are in a theory with three (or more)
separate scales, and it will demonstrate how to count powers of $v$.

\absatz Velocity power counting in NRQCD and identification of the relevant
energy and momentum r\'egimes has proven more difficult than previously
believed. The first attempt by Lepage and co-workers~\cite{CaswellLepage,BBL}
fell shot as working only in the Coulomb gauge and as being complicated and
incomplete. In a recent article, Beneke and Smirnov~\cite{BenekeSmirnov}
pointed out that the much simpler velocity rescaling rules proposed by Luke and
Manohar for Coulomb interactions~\cite{LukeManohar}, and by Grinstein and
Rothstein for bremsstrahlung processes~\cite{GrinsteinRothstein}, as united by
Luke and Savage~\cite{LukeSavage}, and by Labelle's power counting scheme in
time ordered perturbation theory~\cite{Labelle}, do not reproduce the correct
behaviour of the two gluon exchange contribution to Coulomb scattering between
non-relativistic particles near threshold. This has cast some doubt whether
NRQCD, especially in its dimensionally regularised version~\cite{LukeSavage},
can be formulated using a self-consistent low energy Lagrangean. A recent
article~\cite{hgpub3} has resolved this conflict in a toy model, and this
contribution presents the extension to NRQCD.

It is organised as follows: In Sect.\ \ref{philosophy}, the relevant r\'egimes
of NRQCD are identified, extending the formalism of Luke and
Savage~\cite{LukeSavage} by the soft r\'egime of Beneke and
Smirnov~\cite{BenekeSmirnov}. Sect.\ \ref{rescaling} proposes the rescaling
rules necessary for a Lagrangean with manifest velocity power counting and
gives the vertex and loop velocity power counting rules. An example in Sect.\ 
\ref{bsexample} establishes the necessity of the new, soft r\'egime introduced
in Sect.\ \ref{philosophy}: It is essential for the correct reproduction of the
infrared behaviour of QCD in NRQCD scattering amplitudes. Summary and outlook
conclude the article.  There is also some overlap between the topic of this and
Mike Luke's talk at this workshop, although I tried to set slightly different
priorities.

\section{Idea of Dimensionally Regularised NRQCD}
\seclabel{philosophy}

The NRQCD propagators are
\begin{equation}
  \label{nonrelprop}
  Q\;:\;\frac{\ii\;\mathrm{Num}}{T-\frac{\pv^2}{2M}+\ii\epsilon}\;\;,\;\;
   \Amu\;:\;\frac{\ii\;\mathrm{Num}}{k^2+\ii\epsilon}\;\;,
\end{equation}
where $T=p_0-M=\frac{\pv^2}{2M}+\dots$ is the kinetic energy of the quark.
``$\mathrm{Num}$'' are numerators containing the appropriate colour, Dirac and
flavour indices and the gauge fixing term for the gluons, all of which are
unimportant for the considerations in this section.

Cuts and poles in scattering amplitudes close to threshold stem from bound
states and on-shell propagation of particles in intermediate states. They give
rise to infrared divergences, and in general dominate contributions to
scattering amplitudes.  With the two scales at hand, and energies and momenta
being of either scale, three r\'egimes are identified in which either the quark
or the gluon in (\ref{nonrelprop}) is on shell:
\begin{eqnarray}
   \mbox{soft r\'egime: }&\Asmu:&\;k_0\sim |\kv|\sim Mv\;\;,\non\\
  \label{regimes}
   \mbox{potential r\'egime: }&Q_\mathrm{p}:&\;T\sim Mv^2\;,\; |\pv|\sim
   Mv\;\;,\\
   \mbox{ultrasoft r\'egime: }&\Aumu:&\;k_0\sim |\kv|\sim Mv^2\non
\end{eqnarray}
Ultrasoft gluons $\Aumu$ are emitted as bremsstrahlung or from excited states
in the bound system, and hence physical. Soft gluons $\Asmu$ do not describe
bremsstrahlung: Because in- and outgoing quarks $Q_\mathrm{p}$ are close to
their mass shell, they have an energy of order $Mv^2$. Therefore, overall
energy conservation forbids all processes with outgoing soft gluons but without
ingoing ones, and vice versa, as their energy is of order $Mv$.

The list of particles is not yet complete: In a bound system, one needs gluons
which change the quark momenta but keep them close to their mass shell,
relating the (instantaneous) Coulomb interaction:
\begin{equation}
  \label{pgluon}
  \Apmu\;\;:\;\;k_0\sim Mv^2\;,\;|\kv|\sim Mv
\end{equation}
So far, only potential gluons and quarks, and ultrasoft gluons had been
identified in the literature of power counting in
NRQCD~\cite{LukeManohar,GrinsteinRothstein,Labelle}. That the soft r\'egime was
overlooked cast doubts on the completeness of NRQCD after Beneke and
Smirnov~\cite{BenekeSmirnov} demonstrated its relevance near threshold in
explicit one- and two-loop calculations. Here, the fields representing a
non-relativistic quark or gluon came naturally by identifying all possible
particle poles in the non-relativistic propagators, given the two scales at
hand.

When a soft gluon $\Asmu$ couples to a potential quark $ Q_\mathrm{p}$, the
outgoing quark is far off its mass shell and carries energy and momentum of
order $Mv$. Therefore, consistency requires the existence of quarks in the soft
r\'egime as well,
\begin{equation}
  \label{squark}
  Q_\mathrm{s}\;\;:\;\;T\sim |\pv|\sim Mv\;\;.
\end{equation}
As the potential quark has a much smaller energy than either of the soft
particles, it can -- by the uncertainty relation -- not resolve the precise
time at which the soft quark emits or absorbs the soft gluon. So, we expect a
``temporal'' multipole expansion to be associated with this vertex. In general,
the coupling between particles of different r\'egimes will not be point-like
but contain multipole expansions for the particle belonging to the weaker
kinematic r\'egime. For the coupling of potential quarks to ultrasoft gluons,
this has been observed by Grinstein and Rothstein~\cite{GrinsteinRothstein},
and by Labelle~\cite{Labelle}.

Propagators will also be different from r\'egime to r\'egime. In order to
clarify the relation of the presentation here and the work by Beneke and
Smirnov~\cite{BenekeSmirnov}, let us consider a typical loop integral in NRQCD.
One may expand the integrand about the various saddle points, i.e.\ about the
values of the loop-momentum $q$ where particles become on shell. For example,
expanding about a saddle point coming from a physical gluon at the soft scale,
a quark propagator may be expanded as ($T_\mathrm{p}\sim
\frac{\vec{p}^2}{2M}\sim Mv^2\ll q_0\sim|\vec{q}|\sim Mv$)
\begin{equation}
  \label{bsexprop}
   \frac{\ii}{{q_{0,\rm s}}+{T_\mathrm{p}}-\frac{(\vec{p}+\vec{q})^2}{2M}}
   \longrightarrow
   \frac{\ii}{{q_{0,\rm s}}}+\frac{\ii}{{q_{0,\rm s}}}\;\ii\bigg(
     {T_\mathrm{p}}-\frac{(\vec{p}+\vec{q})^2}{2M}\bigg)\;
     \frac{\ii}{{q_{0,\rm s}}}+\dots\;\;.
\end{equation}
So, $\Qs$ is expected to become static to lowest order, and the higher order
terms in the expansion can be interpreted as insertions into the soft quark
propagator, or (as mentioned above and to be confirmed below) as resulting from
an energy multipole expansion which modifies the vertex rules.  As the energy
of potential gluons is much smaller than their momentum, the $\Ap$-propagator
is expected to become instantaneous for similar reasons.

With these five fields $\Qs\;,\Qp\;,\Asmu\;,\Apmu\;,\Aumu$ representing quarks
and gluons in the three different non-relativistic r\'egimes, soft, potential
and ultrasoft, NRQCD becomes self-consistent. An ultrasoft quark (which would
have a static propagator) is not relevant for this paper. It is hence not
considered, as is a fourth (``exceptional'') r\'egime in which momenta are of
the order $Mv^2$ and energies of the order $Mv$ or any r\'egime in which one of
the scales is set by $M$. They do not derive from poles in propagators, and
hence will be relevant only under ``exceptional'' circumstances. A future
publication~\cite{hgpub4} has to prove that the particle content outlined is
not only consistent but complete.

It is worth noticing that the particles of the soft r\'egime can neither be
mimicked by potential gluon exchange, nor by contact terms generated by
integrating out the ultraviolet modes: Fields in the soft r\'egime have momenta
of the same order as the momenta of the potential r\'egime, but much higher
energies. Therefore, seen from the potential scale they describe instantaneous
but non-local interactions, as pointed out by Beneke and
Smirnov~\cite{BenekeSmirnov}. Integrating out the scale $Mv$, one arrives at
soft gluons and quarks as point-like multi-quark interactions in the ultrasoft
r\'egime. The physics of potential quarks and gluons will still have to be
described by spatially local, but non-instantaneous interactions. This suggests
once more that there is no overlap between interactions and particles in
different r\'egimes.

Finally, the regularisation scheme must be chosen such that the three kinematic
r\'egimes still do not overlap, i.e.\ such that expansion around one saddle
point in the loop integral does not obtain any contribution from other saddle
points and r\'egimes.  One might use an energy and momentum cutoff separating
the soft from the potential, and the potential from the ultrasoft r\'egime, but
the integrals encountered can in general not be performed analytically.
Furthermore, cutoff regularisation usually jeopardises power counting and
symmetries, and introduces unphysical power divergences as the (unphysical)
cutoff is removed.  In contradistinction, using dimensional regularisation
\emph{after} the saddle point expansion preserves power counting and gauge
symmetry. Its homogeneity~\cite{BenekeSmirnov} guarantees that contributions
from different saddle points and r\'egimes do not overlap (A simple example can
be found in Ref.\ ~\cite{hgpub3}.). Therefore, dimensional regularisation will
be the method of choice in the example of Sect.\ \ref{bsexample}.

\section{Velocity Power Counting}
\seclabel{rescaling}

\subsection{Rescaling Rules and Propagators}
\label{sec:props}

In order to establish explicit velocity power counting in the NRQCD Lagrangean,
one rescales the space-time coordinates such that typical momenta in either
r\'egime are dimensionless, as proposed by Luke and Manohar~\cite{LukeManohar}
for the potential r\'egime, and by Grinstein and
Rothstein~\cite{GrinsteinRothstein} for the ultrasoft one:
\begin{eqnarray}
  \mbox{soft: } && t=(Mv)^{-1} \;\ts\;\;,\;\; \xv=(Mv)^{-1}\;\xs\;\;,\non\\
  \label{xtscaling}
  \mbox{potential: }&& t=(Mv^2)^{-1}\; \tu\;\;,\;\; \xv=(Mv)^{-1}\;\xs\;\;,\\
\mbox{ultrasoft: }&& t=(Mv^2)^{-1}\; \tu\;\;,\;\;\xv=(Mv^2)^{-1}\;\xu\;\;.\non
\end{eqnarray}
For the propagator terms in the NRQCD Lagrangean to be properly normalised, one
sets for the representatives of the gluons in the three r\'egimes
\begin{eqnarray}
   \mbox{soft: } && \Asmu(\xv,t) = (Mv)\; \calAsmu(\xs,\ts)\;\;,\non\\
   \label{gluonscaling}
   \mbox{potential: } && \Apmu(\xv,t) = (Mv^{\frac{3}{2}})\;
                                           \calApmu(\xs,\tu)\;\;,\\
   \mbox{ultrasoft: } && \Aumu(\xv,t) = (Mv^2)\; \calAumu(\xu,\tu)\;\;,\non
\end{eqnarray}
and for the quark representatives
\begin{eqnarray}
  \label{quarkscaling}
  \mbox{soft: } && Q_\mathrm{s}(\xv,t) = (Mv)^{\frac{3}{2}}\;
  \calQs(\xs,\ts)\;\;,\\
  \mbox{potential: } &&Q_\mathrm{p}(\xv,t) = (Mv)^{\frac{3}{2}}\;
  \calQp(\xs,\tu)\;\;.\non
\end{eqnarray}
The rescaled free quark Lagrangean reads then
\begin{eqnarray}
  \label{qsfreelagr}
   \mbox{soft: } &&\dedreiXs\deTs\calQs^\dagger\Big(\ii\de_0+
   \frac{{v}}{2} \;\vec{\de}^2 \Big)\calQs\;\;,\\
  \label{qpfreelagr}
   \mbox{potential: } &&\dedreiXs\deTu
   \calQp^\dagger\Big(\ii\de_0+\frac{1}{2}\; \vec{\de}^2 \Big)\calQp\;\;.
\end{eqnarray}
Here, as in the following, the positions of the fields have been left out
whenever they coincide with the variables of the volume element.  Derivatives
are to be taken with respect to the rescaled variables of the volume element.

The gauge fixing term was included in the NRQCD Lagrangean (\ref{nrqcdlagr})
because the decomposition of the Lagrangean into a free and an interaction part
is gauge dependent. Usually, the Coulomb gauge $\dev\cdot\Av=0$ is chosen in
NRQCD, but Luke and Savage~\cite{LukeSavage} showed how to establish explicit
velocity power counting in any gauge.  Because of the difference between
canonical and physical momentum, it is important to specify the gauge
\emph{before} identifying to which order in $v$ a certain r\'egime in the
Lagrangean contributes, as seen shortly. Still, the classification of the three
kinematic r\'egimes (\ref{regimes}) itself relied only on the typical
excitation energy and momentum, and hence on gauge invariant quantities, and
the denominator in the gluon propagator (\ref{nonrelprop}) is gauge
independent.

The rescaled free gluon Lagrangean in the Lorentz gauge reads for example
\begin{eqnarray}
  \label{gsfreelagrlorentz}
  \mbox{soft: } &&\dedreiXs\deTs
  \frac{1}{2}\;\calAsmu \Big[\de^2 g_{\mu\nu}-(1-\frac{1}{\alpha})\de_\mu
  \de_\nu\Big]\calA_\mathrm{s}^{\nu}\;\;,\\
  \label{gpfreelagrlorentz}
  \mbox{potential: } &&\dedreiXs\deTu
  \frac{1}{2}\;\calA_{\mathrm{p}}^\mu \Big[ g_{\mu\nu} (v^2\de_0^2-\dev^2)-\\
  &&\phantom{\dedreiXs\deTu\frac{1}{2}\;}-
  (1-\frac{1}{\alpha}) (v \delta_{\mu 0}\de_0+\delta_{\mu i}\de_i)
  (v \delta_{\nu 0}\de_0+\delta_{\nu i}\de_i)\Big]\calA_\mathrm{u}^{\nu}
  \;\;,\non\\
  \label{gufreelagrlorentz}
  \mbox{ultrasoft: } && \dedreiXu\deTu
  \frac{1}{2}\;\calAumu \Big[\de^2 g_{\mu\nu}-(1-\frac{1}{\alpha})\de_\mu
  \de_\nu\Big]\calA_\mathrm{u}^{\nu}\;\;,
\end{eqnarray}
(colour indices suppressed), while in the Coulomb gauge
\begin{eqnarray}
  \label{gsfreelagrcoulomb}
  \mbox{soft: } &&\dedreiXs\deTs
  \half\;{\calA_{i,\mathrm{s}}}
        \Big[(\dev^2-\de_0^2)\delta_{ij}-\de_i\de_j\Big]
      {\calA_{j,\mathrm{s}}}\;\;,\\
  \label{gpfreelagrcoulomb}
  \mbox{potential: } &&\dedreiXs\deTu
   \half\;\Big[{\calA_{0,\mathrm{p}}}\dev^2{\calA_{0,\mathrm{p}}}+
     {\calA_{i,\mathrm{p}}}
        (\dev^2\delta_{ij}-\de_i\de_j-{v^2}\de_0^2\delta_{ij})
      {\calA_{j,\mathrm{p}}}\Big],\\
  \label{gufreelagrcoulomb}
  \mbox{ultrasoft: } &&\dedreiXs\deTs
  \half\;{\calA_{i,\mathrm{u}}}\Big[(\dev^2-\de_0^2)\delta_{ij}-\de_i\de_j\Big]
      {\calA_{j,\mathrm{u}}}\;\;.
\end{eqnarray}
The (un-rescaled) Coulomb gauge propagators are therefore (Dirac and colour
indices suppressed, $\delta_\mathrm{tr}^{ij}=\delta^{ij}-
\frac{k^i k^j}{\kv^2}$)
\begin{eqnarray}
  \label{sprop}
       \mbox{soft: } && \Qs:\;
       \feynbox{60\unitlength}{
       \begin{fmfgraph*}(60,30)
         \fmfleft{i}
         \fmfright{o}
         \fmf{heavy,width=thin,label=$\fs(T,,\pv)$,label.side=left}{i,o}
       \end{fmfgraph*}}
  \;=\;\frac{\ii}{T+\ii\epsilon}
  \;\;,\;\;
  \Avs:\;
    \feynbox{60\unitlength}{
    \begin{fmfgraph*}(60,30)
      \fmfleft{i}
      \fmfv{label=$\fs i$,label.angle=90,label.dist=0.1w}{i}
      \fmfright{o}
      \fmfv{label=$\fs j$,label.angle=90,label.dist=0.1w}{o}
      \fmf{zigzag,width=thin,label=$\fs k$,label.side=left}{i,o}
    \end{fmfgraph*}}
      \;=\;\frac{\ii\;\delta_\mathrm{tr}^{ij}}{k^2+\ii\epsilon}\;\;,
      \\
  \label{pprop}
    \mbox{potential: } && \Qp:\;
    \feynbox{60\unitlength}{
    \begin{fmfgraph*}(60,30)
      \fmfleft{i}
      \fmfright{o}
      \fmf{fermion,width=thick,label=$\fs(T,,\pv)$,label.side=left}{i,o}
    \end{fmfgraph*}}
  \;=\; \frac{\ii}{T-\frac{\pv^2}{2M}+\ii\epsilon}\;\;,\\
  &&
    A_{\mathrm{p},0}:\;
      \feynbox{60\unitlength}{
      \begin{fmfgraph*}(60,30)
        \fmfleft{i}
        \fmfright{o}
        \fmf{dashes,width=thin,label=${\fs k,,A_0}$,label.side=left}{i,o}
      \end{fmfgraph*}}
    \;=\;\frac{-\ii}{-\kv^2+\ii\epsilon}\;\;,\;\;
    \Avp:\;
      \feynbox{60\unitlength}{
      \begin{fmfgraph*}(60,30)
        \fmfleft{i}
        \fmfv{label=$\fs i$,label.angle=90,label.dist=0.1w}{i}
        \fmfright{o}
        \fmfv{label=$\fs j$,label.angle=90,label.dist=0.1w}{o}
        \fmf{dashes,width=thin,label=${\fs k,,\Av}$,label.side=left}{i,o}
      \end{fmfgraph*}}
    \;=\;\frac{\ii\;\delta_\mathrm{tr}^{ij}}{-\kv^2+\ii\epsilon}\;\;,\non\\
    \label{uprop} \mbox{ultrasoft: }
    && \Avu:\;
      \feynbox{60\unitlength}{
      \begin{fmfgraph*}(60,30)
        \fmfleft{i}
        \fmfv{label=$\fs i$,label.angle=90,label.dist=0.1w}{i}
        \fmfright{o}
        \fmfv{label=$\fs j$,label.angle=90,label.dist=0.1w}{o}
        \fmf{photon,width=thin,label=$\fs k$,label.side=left}{i,o}
      \end{fmfgraph*}}
   \;=\;\frac{\ii\;\delta_\mathrm{tr}^{ij}}{k^2+\ii\epsilon}\;\;.
\end{eqnarray}
As expected, the soft quark becomes static and the potential gluon becomes
instantaneous in both gauges. In order to maintain velocity power counting,
corrections of order $v$ or higher must be treated as insertions, represented
in Coulomb gauge by the (un-rescaled) Feynman rules
\begin{equation}
  \label{insertions}
      \feynbox{60\unitlength}{
      \begin{fmfgraph*}(60,30)
        \fmfleft{i}
        \fmfright{o}
        \fmf{double,width=thin}{i,v,o}
        \fmfv{decor.shape=cross,label=$\fs(T,,\pv)$,label.angle=90,label.dist=0.2w}{v}
      \end{fmfgraph*}}
   \;=\;-\ii\;\frac{\pv^2}{2M}\;=\;\calO(v^1)
   \;\;,\;\;
      \feynbox{60\unitlength}{
      \begin{fmfgraph*}(60,30)
        \fmfleft{i}
        \fmfright{o}
        \fmfv{label=$\fs i$,label.angle=90,label.dist=0.1w}{i}
        \fmfv{label=$\fs j$,label.angle=90,label.dist=0.1w}{o}
        \fmf{dashes,width=thin}{i,v,o}
        \fmfv{decor.shape=cross,label=${\fs k,,\Av}$,label.angle=90,label.dist=0.2w}{v}
      \end{fmfgraph*}}
    \;=\;+ \ii k_0^2\;\delta_{ij}\;=\;\calO(v^2)\;\;.
\end{equation}
The Lorentz gauge propagators and insertions are written down
straightforwardly, too~\cite{LukeSavage}, and look very similar to the Coulomb
gauge result, especially for $\alpha=1$. As seen from
(\ref{gsfreelagrcoulomb}--\ref{gufreelagrcoulomb}), the choice of the Coulomb
gauge makes $A_0$ instantaneous, and hence it contributes in the potential
r\'egime, only. Since in this gauge, $A_0$ solely mediates the instantaneous
Coulomb potential (physical fields are transverse by virtue of Gau\3' law),
this result was to be expected. The field $\Avp$ is associated with retardation
effects like spin-orbit coupling and the Darwin term in (\ref{nrqcdlagr}). The
advantages of having couplings between $A_0$ and the other fields only for
potential $A_0$ and of having no insertions in the $A_0$ propagator is balanced
by the demand for non-multiplicative renormalisation of the Coulomb gauge. The
Lorentz gauge may hence facilitate some calculations although the number of
diagrams is larger, as will be seen shortly.

Except for the physical gluons $\Asmu$ and $\Aumu$, there is no distinction
between Feynman and retarded propagators in NRQCD: Antiparticle propagation has
been eliminated by the field transformation from the relativistic to the
non-relativistic Lagrangean, and both propagators have maximal support for
on-shell particles, the Feynman propagator outside the light cone vanishing
like $\e^{-M}$. Feynman's perturbation theory becomes more convenient than the
time-ordered formalism, as less diagrams have to be calculated.

\subsection{Vertex Rules}
\label{sec:vertex}

By experience, particles in the various r\'egimes couple: On-shell (potential)
quarks radiate bremsstrahlung (ultrasoft) gluons. In general, one must allow
all couplings between the various r\'egimes which obey ``scale conservation''
for both energies and momenta. They must be conserved within each r\'egime to
the order in $v$ one works. This will exclude for example the coupling of two
potential quarks ($T\sim Mv^2$) to one soft gluon ($q_0\sim Mv$), but not to
two soft gluons via the $Q^\dagger \vec{A}\cdot\vec{A}Q$ term of the Lagrangean
(\ref{nrqcdlagr}).

As an example, consider a bremsstrahlung-like process: the radiation of a soft
gluon off a soft quark, resulting in a potential quark. The rescaled
interaction Lagrangean reads for the vector coupling
\begin{equation}
  \label{ssplagr}
  \dedreiXs\deTs \Big[-\ii g \;{v}\; \calQs^\dagger(\Xvs,\Ts)\,\dev\cdot
   {\vec{\calAs}}(\Xvs,\Ts) \, \calQp (\Xvs,{v}\Ts)\Big]\;\;.
\end{equation}
Note that the scaling r\'egime of the volume element is set by the particle
with the highest momentum and energy. The Feynman rule for this vertex is hence
\begin{equation}
  \rule{0pt}{26pt}
  \label{sspvertex}
    \feynbox{75\unitlength}{
    \begin{fmfgraph*}(75,30)
      \fmfstraight
      \fmftop{i,o}
      \fmfbottom{u}
      \fmf{heavy,width=thin,label=$\fs(T,,\pv)$,label.side=left,label.dist=0.1w}{i,v}
      \fmf{fermion,width=thick,label=$\fs(T^\prime,,\pv^\prime)$,label.side=right,label.dist=0.1w}{o,v}
      \fmffreeze
      \fmf{zigzag,width=thin,label=$\uparrow\fs q,,i$}{u,v}
    \end{fmfgraph*}}
      \begin{array}{l}
          =-\ii g \;(\pv-\pv^\prime)_i\;(2\pi)^4\;
          \delta^{(3)}(\pv+\pv^\prime+\qv)\times \\
          \;\;\;\;\;\times\Big[\exp\Big({T^\prime\;\frac{\partial}{\partial
          (T+q_0)}}\Big)\;\delta(T+q_0)\Big] \;=\;\calO(v\;\e^v)\;\;.
      \end{array}
\end{equation}
One sees that technically, the energy multipole expansion expected in Sect.\ 
\ref{philosophy} comes from the different scaling of $\xv$ and $t$ in the three
r\'egimes.  The factor $\e^v$ symbolises that the multipole expansion
corresponds term by term to an expansion in $v$. It should be truncated at the
desired order in $v$.

Amongst the fields introduced, six interactions are allowed within and between
the various r\'egimes for the vector coupling, and two (six) for the scalar
coupling in the Coulomb (Lorentz) gauge.  Their $v$ counting is presented in
tables \ref{vectorvertex} and \ref{scalarvertex}.  Note that -- albeit both
describing interactions with physical gluons -- soft and ultrasoft couplings
occur at different orders in $v$, and obey different multipole expansion rules.
On the level of the vertex rules, double counting is prevented by the fact that
in addition to most of the propagators, all vertices are distinct because of
different multipole expansions. Velocity power counting for other vertices is
obtained again by rescaling and multipole expansion. For example, the rules for
the Fermi term $\calL_\mathrm{int}=\frac{g}{2M}\,Q^\dagger
\vec{\sigma}\cdot\vec{B}Q$ are identical to those of table \ref{vectorvertex}.

\begin{table}[!htb]
  \caption{\figlabel{vectorvertex} \sl 
    Velocity power counting and vertices for the
    interaction Lagrangean \protect$-\frac{\ii g}{M}\;Q^\dagger\dev\cdot\Av Q$
    in both Lorentz and Coulomb gauge.} 
\begin{center}
  \footnotesize
\begin{tabular}{|r|c|c|c|c|c|c|}
  \hline
  Vertex\rule[-15pt]{0pt}{35pt}&
        \feynbox{40\unitlength}{
        \begin{fmfgraph*}(40,25)
          \fmfstraight
          \fmftop{i,o}
          \fmfbottom{u}
          \fmf{vanilla,width=thick}{i,v}
          \fmf{vanilla,width=thick}{o,v}
          \fmffreeze
          \fmf{dashes,width=thin,label=$\fs\Av$}{u,v}
        \end{fmfgraph*}}&
      \feynbox{40\unitlength}{
      \begin{fmfgraph*}(40,25)
        \fmfstraight
        \fmftop{i,o}
        \fmfbottom{u}
        \fmf{vanilla,width=thick}{i,v}
        \fmf{vanilla,width=thick}{o,v}
        \fmffreeze
        \fmf{photon,width=thin}{u,v}
      \end{fmfgraph*}}&
      \feynbox{40\unitlength}{
      \begin{fmfgraph*}(40,25)
        \fmfstraight
        \fmftop{i,o}
        \fmfbottom{u}
        \fmf{double,width=thin}{i,v}
        \fmf{double,width=thin}{o,v}
        \fmffreeze
        \fmf{zigzag,width=thin}{u,v}
      \end{fmfgraph*}}&
      \feynbox{40\unitlength}{
      \begin{fmfgraph*}(40,25)
        \fmfstraight
        \fmftop{i,o}
        \fmfbottom{u}
        \fmf{double,width=thin}{i,v}
        \fmf{vanilla,width=thick}{o,v}
        \fmffreeze
        \fmf{zigzag,width=thin}{u,v}
      \end{fmfgraph*}}&
      \feynbox{40\unitlength}{
      \begin{fmfgraph*}(40,25)
        \fmfstraight
        \fmftop{i,o}
        \fmfbottom{u}
        \fmf{double,width=thin}{i,v}
        \fmf{double,width=thin}{o,v}
        \fmffreeze
        \fmf{dashes,width=thin,label=$\fs\Av$}{u,v}
      \end{fmfgraph*}}&
      \feynbox{40\unitlength}{
      \begin{fmfgraph*}(40,25)
        \fmfstraight
        \fmftop{i,o}
        \fmfbottom{u}
        \fmf{double,width=thin}{i,v}
        \fmf{double,width=thin}{o,v}
        \fmffreeze
        \fmf{photon,width=thin}{u,v}
      \end{fmfgraph*}}
  \\[2ex]
  \hline
  \protect$v$ power\rule[-8pt]{0pt}{20pt}&
  \protect$\sqrt{v}$&
  \protect$v$&
  \protect$v$&
  \protect$v$&
  \protect$v^{\frac{3}{2}}$&
  \protect$v^2$
  \\
  \hline
\end{tabular}
\end{center}
\end{table}
\vspace*{-5ex} 
\begin{table}[!htb]
  \caption{\figlabel{scalarvertex} \sl 
     Velocity power counting and vertices for the
     interaction Lagrangean \protect$-g Q^\dagger A_0 Q$ in the Lorentz gauge.
     In the Coulomb gauge, only the first two diagrams exist.}
\begin{center}
  \footnotesize
\begin{tabular}{|r|c|c|c|c|c|c|}
  \hline
  Vertex\rule[-15pt]{0pt}{35pt}&
      \feynbox{40\unitlength}{
      \begin{fmfgraph*}(40,25)
        \fmfstraight
        \fmftop{i,o}
        \fmfbottom{u}
        \fmf{vanilla,width=thick}{i,v}
        \fmf{vanilla,width=thick}{o,v}
        \fmffreeze
        \fmf{dashes,width=thin,label=$\fs A_0$}{u,v}
      \end{fmfgraph*}}&
      \feynbox{40\unitlength}{
        \begin{fmfgraph*}(40,25)
        \fmfstraight
        \fmftop{i,o}
        \fmfbottom{u}
        \fmf{double,width=thin}{i,v}
        \fmf{double,width=thin}{o,v}
        \fmffreeze
        \fmf{dashes,width=thin,label=$\fs A_0$}{u,v}
      \end{fmfgraph*}}&
      \feynbox{40\unitlength}{
      \begin{fmfgraph*}(40,25)
        \fmfstraight
        \fmftop{i,o}
        \fmfbottom{u}
        \fmf{vanilla,width=thick}{i,v}
        \fmf{vanilla,width=thick}{o,v}
        \fmffreeze
        \fmf{photon,width=thin}{u,v}
      \end{fmfgraph*}}&
      \feynbox{40\unitlength}{
      \begin{fmfgraph*}(40,25)
        \fmfstraight
        \fmftop{i,o}
        \fmfbottom{u}
        \fmf{double,width=thin}{i,v}
        \fmf{double,width=thin}{o,v}
        \fmffreeze
        \fmf{zigzag,width=thin}{u,v}
      \end{fmfgraph*}}&
      \feynbox{40\unitlength}{
      \begin{fmfgraph*}(40,25)
        \fmfstraight
        \fmftop{i,o}
        \fmfbottom{u}
        \fmf{double,width=thin}{i,v}
        \fmf{vanilla,width=thick}{o,v}
        \fmffreeze
        \fmf{zigzag,width=thin}{u,v}
      \end{fmfgraph*}}&
      \feynbox{40\unitlength}{
      \begin{fmfgraph*}(40,25)
        \fmfstraight
        \fmftop{i,o}
        \fmfbottom{u}
        \fmf{double,width=thin}{i,v}
        \fmf{double,width=thin}{o,v}
        \fmffreeze
        \fmf{photon,width=thin}{u,v}
      \end{fmfgraph*}}
  \\[2ex]
  \hline
  \protect$v$ power\rule[-8pt]{0pt}{20pt}&
  \protect$\frac{1}{\sqrt{v}}$&
  \protect$\sqrt{v}$&
  \protect$v^0$&
  \protect$v^0$&
  \protect$v^0$&
  \protect$v$
  \\
  \hline
\end{tabular}
\end{center}
\end{table}

Using the equations of motion, a temporal multipole expansion may be re-written
such that the energy becomes conserved at the vertex. Now, both soft and
potential or ultrasoft energies are present in the propagators, making it
necessary to expand it in ultrasoft and potential energies. An example would be
to restate the vertex (\ref{sspvertex}) as
\begin{eqnarray}
    \feynbox{40\unitlength}{
    \begin{fmfgraph*}(40,20)
      \fmfstraight
      \fmftop{i,o}
      \fmfbottom{u}
      \fmf{double,width=thin}{i,v}
      \fmf{vanilla,width=thick}{o,v}
      \fmffreeze
      \fmf{zigzag,width=thin}{u,v}
    \end{fmfgraph*}}
   &=&-\ii g\;(\pv-\pv^\prime)_i\;(2\pi)^4\;\delta(T+T^\prime+q_0)
   \;\delta^{(3)}(\pv+\pv^\prime+\qv)\;\;,
\end{eqnarray}
and the soft propagator to contain insertions $\calO(v)$ for potential energies
$T^\prime$
\begin{equation}
    \feynbox{60\unitlength}{
    \begin{fmfgraph*}(60,30)
      \fmfleft{i}
      \fmfright{o}
      \fmf{heavy,width=thin,label=$\fs(-T=q_0+T^\prime,,\pv)$,label.side=left}{i,o}
      \end{fmfgraph*}}
    \;=\;\frac{\ii}{q_0+\ii\epsilon}
    \sum\limits_{n=0}^\infty\left(\frac{-T^\prime}{q_0}\right)^n \;\;.
\end{equation}
This gives us the first part of the correspondence to Beneke and Smirnov's
threshold expansion in the example propagator (\ref{bsexprop}), the second one
coming from the soft quark insertion (\ref{insertions}).  The same can be shown
for the momentum-non-conserving vertices, too.

It is also interesting to note that there is no choice but to assign one and
the same coupling strength $g$ to each interaction. Different couplings for one
vertex in different r\'egimes are not allowed. This is to be expected, as the
connection to Beneke and Smirnov's threshold expansion~\cite{BenekeSmirnov}
demonstrated that the fields in the various r\'egimes are representatives of
one and the same non-relativistic particle, whose interactions are fixed by the
non-relativistic Lagrangean (\ref{nrlagr}).
  
In the renormalisation group approach, there is only one relevant coupling
(i.e.\ only one which dominates at zero velocity): As expected, it is the
$\Qp\Qp A_{\mathrm{p},0}$ coupling providing the binding. In the Coulomb gauge,
all other couplings and insertions are irrelevant, while the Lorentz gauge
exhibits three marginal couplings: $\Qp\Qp A_{\mathrm{u},0}\;,\Qs\Qs
A_{\mathrm{s},0}$ and $\Qs\Qp A_{\mathrm{s},0}$.

\subsection{Loop Rules}
\label{sec:loop}

The velocity power counting is not yet complete. As one sees from the volume
element used in (\ref{ssplagr}), the vertex rules for the soft r\'egime count
powers of $v$ with respect to the soft r\'egime. One hence retrieves the
velocity power counting of Heavy Quark Effective
Theory~\cite{IsgurWise1,IsgurWise2} (HQET), in which the interactions between
one heavy (and hence static) and one or several light quarks are
described\footnote{Usually, HQET counts inverse powers of mass in the
  Lagrangean, but because in the soft r\'egime $Mv\sim \mbox{const.}$, the two
  approaches are actually equivalent.}.  HQET becomes a sub-set of NRQCD,
complemented by interactions between soft (HQET) and potential or ultrasoft
particles.

In NRQCD with two potential quarks as initial and final states, the soft
r\'egime can occur only inside loops, as noted above. Therefore, the power
counting in the soft sub-graph has to be transfered to the potential r\'egime.
Because soft loop momenta scale like $[\dd^{4}\! k_\mathrm{s}]\sim v^4$, while
potential ones like $[\dd^{4}\! k_\mathrm{p}]\sim v^5$, each largest sub-graph
which contains only soft quarks and no potential ones (a ``soft blob'') is
enhanced by an additional factor $\frac{1}{v}$.

As an example, consider the graphs of Fig.\ \ref{loopfigs}: Using the Lorentz
gauge, vertex power counting gives that the leading contribution is from the
exchange of two potential gluons, coupled via $Q^\dagger Q A_0$. There are four
such vertices, so the diagram is $\calO(g^4v^{-2})$ (table \ref{scalarvertex}).
The next two diagrams are $\calO(g^4v^0)$ and $\calO(g^6v^0)$ from the vertex
power counting, but another factor $\frac{1}{v}$ must be included because there
is one soft blob in the diagrams. The intermediate couplings in the third
diagram take place in the soft r\'egime and hence are counted in that r\'egime.
The last diagram, in which two soft blobs are separated by the propagation of
two potential quarks is $\calO(g^8 v^0)$ from the vertices, and the loop
counting gives a factor $\frac{1}{v^2}$. Each soft blob contributes at
least four orders of $g$, but only one inverse power of $v\sim g^2$. Power
counting is preserved.  These velocity power counting rules in loops are
verified in explicit calculations of the exemplary graphs (see also below), but
a rigorous derivation is left for a future publication~\cite{hgpub4}.

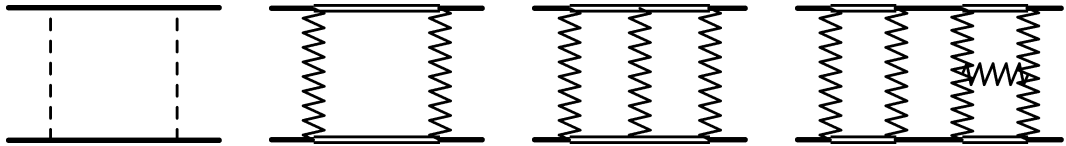
\begin{figure}[!htb]
\begin{center}
  \feynbox{80\unitlength}{
  \begin{fmfgraph*}(80,50)
    \fmfstraight
    \fmftop{i1,o1}
    \fmfbottom{i2,o2}
    \fmf{vanilla,width=thick,tension=1.5}{i1,v1}
    \fmf{vanilla,width=thick,tension=0.5}{v1,v2}
    \fmf{vanilla,width=thick,tension=1.5}{v2,o1}
    \fmf{vanilla,width=thick,tension=1.5}{i2,v3}
    \fmf{vanilla,width=thick,tension=0.5}{v3,v4}
    \fmf{vanilla,width=thick,tension=1.5}{v4,o2}
    \fmffreeze
    \fmf{dashes,width=thin,tension=0.5}{v3,v1}
    \fmf{dashes,width=thin,tension=0.5}{v4,v2}
  \end{fmfgraph*}}
\hspace*{1em}
  \feynbox{80\unitlength}{
  \begin{fmfgraph*}(80,50)
    \fmfstraight
    \fmftop{i1,o1}
    \fmfbottom{i2,o2}
    \fmf{vanilla,width=thick,tension=1.5}{i1,v1}
    \fmf{double,width=thin,tension=0.5}{v1,v2}
    \fmf{vanilla,width=thick,tension=1.5}{v2,o1}
    \fmf{vanilla,width=thick,tension=1.5}{i2,v3}
    \fmf{double,width=thin,tension=0.5}{v3,v4}
    \fmf{vanilla,width=thick,tension=1.5}{v4,o2}
    \fmffreeze
    \fmf{zigzag,width=thin,tension=0.5}{v3,v1}
    \fmf{zigzag,width=thin,tension=0.5}{v4,v2}
  \end{fmfgraph*}}
\hspace*{1em}
  \feynbox{80\unitlength}{
  \begin{fmfgraph*}(80,50)
    \fmfstraight
    \fmftop{i1,o1}
    \fmfbottom{i2,o2}
    \fmf{vanilla,width=thick,tension=1}{i1,v1}
    \fmf{double,width=thin,tension=0.5}{v1,v5,v2}
    \fmf{vanilla,width=thick,tension=1}{v2,o1}
    \fmf{vanilla,width=thick,tension=1}{i2,v3}
    \fmf{double,width=thin,tension=0.5}{v3,v6,v4}
    \fmf{vanilla,width=thick,tension=1}{v4,o2}
    \fmffreeze
    \fmf{zigzag,width=thin,tension=0.5}{v3,v1}
    \fmf{zigzag,width=thin,tension=0.5}{v4,v2}
    \fmf{zigzag,width=thin,tension=0.5}{v5,v6}
  \end{fmfgraph*}}
\hspace*{1em}
  \feynbox{100\unitlength}{
  \begin{fmfgraph*}(100,50)
    \fmfstraight
    \fmftop{i1,o1}
    \fmfbottom{i2,o2}
    \fmf{vanilla,width=thick,tension=1}{i1,v1}
    \fmf{double,width=thin,tension=0.5}{v1,v5}
    \fmf{double,width=thin,tension=0.5}{v7,v2}
    \fmf{vanilla,width=thick,tension=1}{v2,o1}
    \fmf{vanilla,width=thick,tension=1}{i2,v3}
    \fmf{double,width=thin,tension=0.5}{v3,v6}
    \fmf{double,width=thin,tension=0.5}{v8,v4}
    \fmf{vanilla,width=thick,tension=1}{v4,o2}
    \fmf{vanilla,width=thick,tension=0.5}{v5,v7}
    \fmf{vanilla,width=thick,tension=0.5}{v6,v8}
    \fmffreeze
    \fmf{zigzag,width=thin,tension=0.5}{v3,v1}
    \fmf{zigzag,width=thin,tension=0.5}{v4,v9,v2}
    \fmf{zigzag,width=thin,tension=0.5}{v5,v6}
    \fmf{zigzag,width=thin,tension=0.5}{v7,v10,v8}
    \fmffreeze
    \fmf{zigzag,width=thin,tension=0.5}{v9,v10}
  \end{fmfgraph*}}
\end{center}
\caption{\figlabel{loopfigs} Power counting with soft loops. The loops in the
  second and third diagram obtain an inverse power of \protect$v$, the last
  diagram of \protect$v^2$ in addition to the power counting following from the
  vertex rules (\protect\ref{scalarvertex}).}
\end{figure}

With rescaling, multipole expansion and loop counting, the velocity power
counting rules are established, and one can now proceed to check the validity
of the proposed Lagrangean by matching NRQCD to the relativistic theory in an
example given by Beneke and Smirnov~\cite{BenekeSmirnov}.

\section{A Model Calculation}
\seclabel{bsexample}

By construction, NRQCD and QCD must agree in the infrared limit, and especially
in the structure of collinear (infrared) divergences. Matching NRQCD to the low
velocity limit of QCD will therefore confirm that the power counting proposed
is correct and that the soft r\'egime is relevant.
\begin{figure}[!htb]
  \begin{center}
    \feynbox{100\unitlength}{
    \begin{fmfgraph*}(100,60)
       \fmfstraight
       \fmftop{i1,o1}
       \fmfbottom{i2,o2}
       \fmf{fermion,width=thick}{i1,v1,o1}
       \fmf{fermion,width=thick}{i2,v2,o2}
       \fmffreeze
       \fmf{zigzag,left=0.5,width=thin}{v1,v2,v1}
    \end{fmfgraph*}}
  \hspace*{5em}
    \feynbox{100\unitlength}{
    \begin{fmfgraph*}(100,60)
       \fmfstraight
       \fmftop{i1,o1}
       \fmfbottom{i2,o2}
       \fmf{fermion,width=thick}{i1,v1,v2,o1}
       \fmf{fermion,width=thick}{i2,v3,v4,o2}
       \fmffreeze
       \fmf{dashes,width=thin,label=$\fs A_0$}{v1,v3}
       \fmf{dashes,width=thin,label=$\fs A_0$}{v4,v2}
    \end{fmfgraph*}} 
  \end{center}
\caption{\figlabel{graph1}\sl Leading soft (left) and potential (right) planar
  contribution to quark-quark scattering in the Coulomb gauge.}
\end{figure}
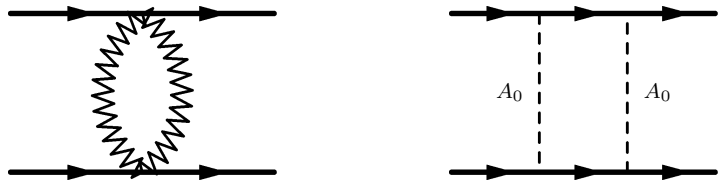
In QCD, the lowest order infrared divergent contribution in the Coulomb gauge
which comes from the soft r\'egime is the one loop graph Fig.\ \ref{graph1} of
order $g^4 v$, while the leading contribution at this order in $g$ comes from
the two Coulomb gluon exchange and is order $v^{-2}$. The leading diagram
mediating retardation effects ($\Avp$ couplings) in this order in $g$ is
$\calO(v^2)$. At high enough order in $g$, soft physical gluon exchange will
therefore be more significant than retardation effects. The number of diagrams
to be considered in the case at hand is large enough\footnote{Actually, most of
  them are zero, but to establish a pattern which convinces one of that is not
  the topic of this contribution~\cite{hgpub4}.}
to make one long for a simpler example.

For the sake of simplicity, let us -- following Beneke and
Smirnov~\cite{BenekeSmirnov} -- deal with a toy model NRQFT Lagrangean
\begin{equation}
  \label{nrlagr}
  \calL_\mathrm{NRQFT}=\Phi^\dagger\Big(\ii\de_0+\frac{\dev^2}{2M} - g c_1\;A
  \Big)\Phi + \half(\de_\mu A)(\de^\mu A) + c_2\Big(\Phi^\dagger\Phi\Big)^2 +
  \dots\;\;
\end{equation}
of a heavy, complex scalar field $\Phi$ with mass $M$ coupled to a massless,
real scalar $ A $. The coupling constant $g$ has been chosen dimensionless.
In a slight abuse of language, $\Phi$ will still be referred to as ``quark''
and $A$ as ``gluon''.  The coefficients $c_i$ are again to be determined by
matching relativistic and non-relativistic scattering amplitudes.  This
Lagrangean is very similar to the $A_0$-part of the NRQCD Lagrangean in Lorentz
gauge. Especially, the vertex and loop power counting is identical to the one
of table \ref{scalarvertex}.

The first soft non-zero contribution in this toy model comes from the two gluon
direct exchange diagram of Fig.\ \ref{graph2} calculated by Beneke and
Smirnov~\cite{BenekeSmirnov} using threshold expansion.  The Mandelstam
variable $t=-(\pv-\pv^\prime)^2$ describes the momentum transfer in the center
of mass system, $y=-(\pv)^2\propto - v^2$ the relative four-momentum squared of
the ingoing quarks as indicator for the thresholdness of the process.
\begin{figure}[!htb]
  \vspace*{2ex}
  \begin{center}
    \feynbox{130\unitlength}{
    \begin{fmfgraph*}(100,60)
      \fmfstraight
      \fmftop{i1,o1}
      \fmfbottom{i2,o2}
      \fmf{fermion,width=thick,tension=1,label=$\fs(T,,\pv)$,label.side=left}{i1,v1}
      \fmf{fermion,width=thick,tension=0.5}{v1,v2}
      \fmf{fermion,width=thick,tension=1,label=$\fs(T,,\pv^\prime)$,label.side=left}{v2,o1}
      \fmf{fermion,width=thick,tension=1,label=$\fs(T,,-\pv)$,label.side=right}{i2,v3}
      \fmf{fermion,width=thick,tension=0.5}{v3,v4}
      \fmf{fermion,width=thick,tension=1,label=$\fs(T,,-\pv^\prime)$,label.side=right}{v4,o2}
      \fmffreeze
      \fmf{gluon,width=thin,label=$\fs k{\dis\uparrow}$,label.side=left,tension=0.5}{v3,v1}
      \fmf{gluon,width=thin,label=${\dis\downarrow}{(k_0,,\atop\pv-\pv^\prime+\kv)}$,label.side=right,tension=0.5}{v4,v2}
    \end{fmfgraph*} }
  \bf{=}\hq
    \feynbox{100\unitlength}{
    \begin{fmfgraph*}(100,60)
      \fmfstraight
      \fmftop{i1,o1}
      \fmfbottom{i2,o2}
      \fmf{fermion,width=thick,tension=1.5}{i1,v1}
      \fmf{fermion,width=thick,tension=0.5}{v1,v2}
      \fmf{fermion,width=thick,tension=1.5}{v2,o1}
      \fmf{fermion,width=thick,tension=1.5}{i2,v3}
      \fmf{fermion,width=thick,tension=0.5}{v3,v4}
      \fmf{fermion,width=thick,tension=1.5}{v4,o2}
      \fmffreeze
      \fmf{photon,width=thin,label=$\fs k{\dis\uparrow}$,label.side=left,tension=0.5}{v3,v1}
      \fmf{photon,width=thin,label=$\fs{\dis\downarrow} k$,label.side=right,tension=0.5}{v4,v2}
    \end{fmfgraph*}}
  \hq\bf{+}\hq
    \feynbox{115\unitlength}{
    \begin{fmfgraph*}(100,60)
      \fmfstraight
      \fmftop{i1,o1}
      \fmfbottom{i2,o2}
      \fmf{fermion,width=thick,tension=1.5}{i1,v1}
      \fmf{fermion,width=thick,tension=0.5}{v1,v2}
      \fmf{fermion,width=thick,tension=1.5}{v2,o1}
      \fmf{fermion,width=thick,tension=1.5}{i2,v3}
      \fmf{fermion,width=thick,tension=0.5}{v3,v4}
      \fmf{fermion,width=thick,tension=1.5}{v4,o2}
      \fmffreeze
      \fmf{photon,width=thin,label=$\fs k{\dis\uparrow}$,label.side=left,tension=0.5}{v3,v1}
      \fmf{dashes,width=thin,label=${\dis\downarrow} {(k_0,,\atop\pv^\prime-\pv)} $,label.side=right,tension=0.5}{v4,v2}
    \end{fmfgraph*} }
  \hq\bf{+}\hq

\vspace*{30\unitlength}

\bf{+}\hq
    \feynbox{100\unitlength}{
    \begin{fmfgraph*}(100,60)
      \fmfstraight
      \fmftop{i1,o1}
      \fmfbottom{i2,o2}
      \fmf{fermion,width=thick,tension=1.5}{i1,v1}
      \fmf{fermion,width=thick,tension=0.5}{v1,v2}
      \fmf{fermion,width=thick,tension=1.5}{v2,o1}
      \fmf{fermion,width=thick,tension=1.5}{i2,v3}
      \fmf{fermion,width=thick,tension=0.5}{v3,v4}
      \fmf{fermion,width=thick,tension=1.5}{v4,o2}
      \fmffreeze
      \fmf{dashes,width=thin,tension=0.5}{v3,v1}
      \fmf{photon,width=thin,tension=0.5}{v4,v2}
    \end{fmfgraph*}}
  \hq\bf{+}\hq
    \feynbox{100\unitlength}{
    \begin{fmfgraph*}(100,60)
      \fmfstraight
      \fmftop{i1,o1}
      \fmfbottom{i2,o2}
      \fmf{fermion,width=thick,tension=1.5}{i1,v1}
      \fmf{fermion,width=thick,tension=0.5}{v1,v2}
      \fmf{fermion,width=thick,tension=1.5}{v2,o1}
      \fmf{fermion,width=thick,tension=1.5}{i2,v3}
      \fmf{fermion,width=thick,tension=0.5}{v3,v4}
      \fmf{fermion,width=thick,tension=1.5}{v4,o2}
      \fmffreeze
      \fmf{dashes,width=thin,tension=0.5}{v3,v1}
      \fmf{dashes,width=thin,tension=0.5}{v4,v2}
    \end{fmfgraph*}}
  \hq\bf{+}\hq
    \feynbox{100\unitlength}{
    \begin{fmfgraph*}(100,60)
      \fmfstraight
      \fmftop{i1,o1}
      \fmfbottom{i2,o2}
      \fmf{fermion,width=thick,tension=1.5}{i1,v1}
      \fmf{heavy,width=thin,tension=0.5,label=$\fs(k_0,,\pv+\kv)$,label.side=left}{v1,v2}
      \fmf{fermion,width=thick,tension=1.5}{v2,o1}
      \fmf{fermion,width=thick,tension=1.5}{i2,v3}
      \fmf{heavy,width=thin,tension=0.5,label=$\fs (-k_0,,-\pv-\kv)$,label.side=right}{v3,v4}
      \fmf{fermion,width=thick,tension=1.5}{v4,o2}
      \fmffreeze
      \fmf{zigzag,width=thin,tension=0.5}{v3,v1}
      \fmf{zigzag,width=thin,tension=0.5}{v4,v2}
  \end{fmfgraph*} }
\end{center}
\vspace*{8pt}
\caption{\figlabel{graph2}\sl Planar \protect$\calO(g^4)$ contributions to
  Coulomb scattering in the toy model. The four-point interaction and insertion
  diagrams are not displayed.}
\end{figure}
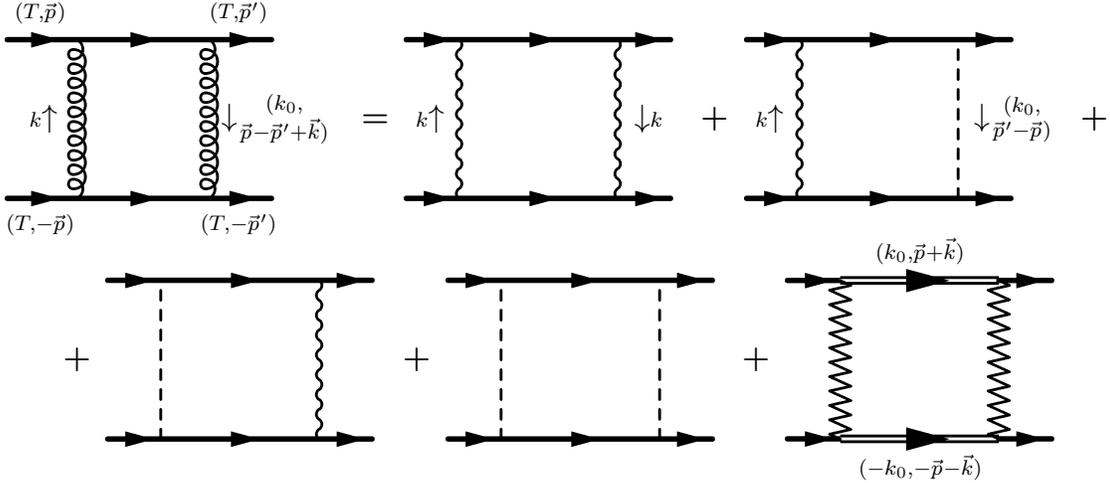
The ultraviolet behaviour of this graph is mimicked in NRQFT by the
four-fermion exchange with $\ii c_2=\frac{-\ii g^4}{24 \pi^2
  M^2}=\calO(t^0,y^0)$, which using the rescaling rules is seen to be
$\calO(v^1)$.

The $\Au\Au$-diagram is of order $\e^v$ (see table \ref{scalarvertex}) with a
leading loop integral contribution (similar to Beneke and
Smirnov's~\cite{BenekeSmirnov} fl.\ (32))
\begin{equation}
  \label{vertex2uu}
  \int\dedk \frac{1}{k_0^2-\kv^2}\;
  \frac{1}{k_0^2-\kv^2}\;\frac{1}{T+k_0-\frac{\pv^2}{2M}}
  \;\frac{1}{T-k_0-\frac{\pv^2}{2M}} \;\;.
\end{equation}
The diagram is expected to be zero since the ultrasoft gluons do not change the
quark momenta and therefore the scattering takes place only in the forward
direction, $\pv=\pv^\prime$.  As no scale is present, it indeed vanishes upon
employing the on-shell condition for potential quarks, $T=\frac{\pv^2}{2M}$ to
leading order\footnote{Since $T-\frac {\pv^2}{2M}\sim Mv^4\ll k\sim Mv\;(Mv^2)$
  in the potential r\'egime, this is a legitimate expansion.}. The $\Au\Ap$
and $\Ap\Au$ contributions ($\calO(\frac{1}{v}\e^v)$) are zero for the same
reason. The lowest order contribution to the $\Ap\Ap$ graph
($\calO(\frac{1}{v^2})$) is:
\begin{equation}
  \label{vertex2pp}
  \int\dedk \frac{1}{\kv^2-\ii\epsilon}\;\frac{1}{(\pv-\pv^\prime+\kv)^2
  -\ii\epsilon}\; \frac{1}{T+k_0-\frac{(\kv+\pv)^2}{2M}+\ii\epsilon}\;
  \frac{1}{T-k_0-\frac{(\kv+\pv)^2}{2M}+\ii\epsilon}
\end{equation}
In the light of the discussion at the end of Sect.\ \ref{philosophy}, it is
most consistent to perform the $k_0$ integration by dimensional regularisation,
using~\cite{Collins} $\int\deintdim{d}{k}=\int\deintdim{\sigma}{k_0}
\deintdim{d-\sigma}{\kv}$, $\sigma\to1$. Split dimensional regularisation was
introduced by Leibbrandt and Williams~\cite{LeibbrandtWilliams} to cure the
problems arising from pinch singularities in non-covariant gauges. Here, it has
the same effect as closing the $k_0$-contour and picking up the quark
propagator poles prior to using dimensional regularisation in $d-1$ Euclidean
dimensions.  Considering also one insertion (\ref{insertions}) at the potential
gluon lines to achieve $\calO(v^1)$ accuracy, the result,
\begin{equation}
  \label{vertex2ppresult}
  \frac{\ii}{8\pi t}\;\frac{M+T}{\sqrt{y}}\;\Big(\frac{2}{4-d}-
  \gamma_\mathrm{E}-\ln\frac{-t}{4\pi\mu^2}\Big)\;\;,
\end{equation}
agrees with Beneke and Smirnov's~\cite{BenekeSmirnov} fl.\ (31) when one keeps
in mind that non-relativistic external lines were normalised differently, and
that different conventions for dimensionally regularised integrals were chosen.
Near threshold, the scale is set by the total threshold energy
$4\pi\mu^2=4(M+T)^2$.

The soft gluon part is to lowest order ($\calO(v^{-1})$) given by
\begin{equation}
  \label{vertex2ss}
  \int\dedk\frac{1}{k_0^2-\kv^2+\ii\epsilon}\;\frac{1}{k_0^2-(\pv-\pv^\prime+
\kv)^2+\ii\epsilon}\;\frac{1}{k_0+\ii\epsilon}\;\frac{1}{-k_0+\ii\epsilon}\;\;,
\end{equation}
which corresponds to Beneke and Smirnov's~\cite{BenekeSmirnov} fl.\ (33). Now,
split dimensional regularisation must be used if no ad-hoc prescription for the
pinch singularity at $k_0=0$ is to be invoked. That the pinch is accounted for
by potential gluon exchange and hence must be discarded, agrees with the
intuitive argument that zero four-momentum scattering in QED is mediated by a
potential only, and no retardation or radiation effects occur. On the other
hand, the model Lagrangean contains three marginal couplings as seen at the end
of Sect.\ \ref{sec:vertex}, which may give finite contributions as energies and
momenta of the scattered particles go to zero. The result to $\calO(v^1)$
exhibits another collinear divergence,
\begin{equation}
  \label{vertex2ssresult}
  \frac{-\ii}{4\pi^2 t}\;\Big(\frac{2}{4-d}-\gamma_\mathrm{E}-
  \ln\frac{-t}{4\pi\mu^2}\Big)\;+\;\frac{\ii}{24\pi^2M^2}\;\Big[1+ \frac{2y}{t}
   \;\Big(\frac{2}{4-d}-\gamma_\mathrm{E}-\ln\frac{-t}{4\pi\mu^2}\Big)\Big]
   \;\;,
\end{equation}
and agrees with fl.\ (36) given by Beneke and Smirnov~\cite{BenekeSmirnov}. The
second term comes from insertions and multipole expansions to achieve
$\calO(v^1)$ accuracy.

It is easy to see that the power counting proposed works.  As expected, the
potential diagram is $\sqrt{y}\propto v$ stronger that the leading soft
contribution, and $t \sqrt{y}\propto v^3$ stronger than the four-fermion
interaction.

In conclusion, the proposed NRQCD power counting and Lagrangean with three
different kinematic r\'egimes (\ref{regimes}) reproduces the collinear
divergences of the planar graph of the relativistic theory \emph{only if} the
soft gluon and the soft quark are accounted for: The four-fermion contact
interaction produces just a $\frac{1}{M^2}$-term, graphs containing ultrasoft
gluons were absent, and the potential gluon (\ref{vertex2ppresult}) gave no
$\calO(y^0)$ contribution. The coupling strength of the $\Phis\As\Phip$ vertex
is also seen to be identical to the other vertex coupling strengths, $g$.

\section{Conclusions and Outlook}
\seclabel{conclusions}

The objective of this contribution was a presentation of the ideas behind
explicit velocity power counting in dimensionally regularised NRQCD. The
identification of three different r\'egimes of scale for on-shell particles in
NRQCD leads in a natural way to the existence of a new quark field and a new
gluon field in the soft scaling r\'egime $E\sim |\pv|\sim Mv$. In it, quarks
are static and gluons on shell, and HQET becomes a sub-set of NRQCD. Neither of
the five fields in the three r\'egimes should be thought of as ``physical
particles''. Rather, they represent the ``true'' quark and gluon in the
respective r\'egimes as the infrared-relevant degrees of freedom. None of the
r\'egimes overlap. An NRQCD Lagrangean has been proposed which leads to the
correct behaviour of scattering and production amplitudes.  It establishes
explicit velocity power counting which is preserved to all orders in
perturbation theory, once dimensional regularisation is chosen to complete the
theory. The reason is non-commutativity of the expansion in small parameters
with dimensionally regularised integrals.

I would like to stress that the diagrammatic threshold expansion derived here
allows for a more automatic and intuitive approach and makes it easier to
determine the order in $v$ to which a certain graph contributes than Beneke and
Smirnov's way~\cite{BenekeSmirnov}. Also, the NRQCD Lagrangean can easily be
applied to bound state problems. An investigation of the influence of soft
quarks and gluons on bound state calculations in NRQED and NRQCD is important
because -- as seen at the beginning of Sect.\ \ref{bsexample}, their
contribution at $\calO(g^4)$ and higher becomes stronger than retardation
effects. As the threshold expansion of Beneke and Smirnov starts in a
relativistic setting, it may formally be harder to treat bound states there.

Coming back to the topic of this workshop, effective nuclear theories, NRQCD
shows how to establish a power counting in any effective field theory with
several low energy scales: First, identify the combinations of scales in which
particles become on shell by looking at the denominators of the various
propagators. This gives the scaling r\'egimes. Then, the Lagrangean is rescaled
to dimensionless fields in each r\'egime to exhibit the vertex and loop power
counting rules. A priori, all couplings obeying scale conservation are allowed.

The problem with effective $NN$ scattering is not that the three scales
$M_N\;,\sqrt{M_N m_\pi}\;,m_\pi$ are separated only by powers of
$\sqrt{\frac{M_N}{ m_\pi}}\approx 0.4$. Indeed, Kaplan, Savage and
Wise~\cite{KSW1,KSW2} obtain very promising results for scales much smaller
than the pion mass (See also David Kaplan's talk in this workshop, and Martin
Savage's contribution on the inclusion of what in the language of this article
would be ultrasoft pions.). The difficulty is that the scale $\sqrt{M_N
  m_\pi}\approx 360 \mathrm{MeV}$ at which the soft r\'egime becomes relevant
is larger than the $NN$ scattering expansion parameter $\Lambda_{NN}\approx
300\mathrm{MeV}$. How to overcome this is another interesting topic for the
future.

\section*{Acknowledgments} 
It is my pleasure express my gratitude to J.-W.\ Chen, D.\ B.\ Kaplan, M.\ Luke
and M.\ J.\ Savage for stimulating discussions. Cordial thanks also to the
organisers and participants of this vivid workshop. The work was supported
in part by a Department of Energy grant DE-FG03-97ER41014.


\end{fmffile}
\end{document}